\documentclass[]{aiaa-tc}
\usepackage{lettrine}
\usepackage{iepc2025}
\usepackage{footmisc}

\usepackage{url}
\usepackage[numbers,sort&compress]{natbib} 
\setcitestyle{open={},close={}} 
\usepackage{multicol}
\usepackage{amsmath}
\usepackage{booktabs, threeparttable, stackengine}
\usepackage{multirow}
\usepackage{subfig}   
\usepackage{xcolor}
\setcitestyle{super} 
\IEPCsubmissionnumber{288} 

\title{Numerical Analysis of Ground Testing for the Intake Device
of an Atmosphere-Breathing Electric Propulsion}

\author{
Geonwoong Moon\thanks{Ph.D. Student, Department of Aerospace Engineering, kw4046@kaist.ac.kr}~%
~and Eunji Jun\thanks{Associate Professor, Department of Aerospace Engineering, eunji.jun@kaist.ac.kr}\\%
{\normalsize\it{Korea Advanced Institute of Science and Technology, Daejeon, 34141, Republic of Korea}}\\%
\\%
Minwoo Yi\thanks{Research Engineer, Satellite Systems PMO, minwooyi@add.re.kr}~,~ Hyunjin Choi\thanks{Research Engineer, Satellite Systems PMO, hchoi1add@add.re.kr}~,~and Kangmin Park\thanks{Research Engineer, Satellite Systems PMO, kmp2002@add.re.kr}\\%
{\normalsize\it{Agency for Defense Development, Daejeon, 34186, Republic of Korea}}\\%
\\%
Younho Kim\thanks{Research Engineer, Space Technology Research Center, yhk@satreci.com}~,~ Jaecheong Lee\thanks{Research Engineer, Space Technology Research Center, jclee@satreci.com}~,~ Jeongjae Lee\thanks{Research Engineer, Space Technology Research Center, jjlee@satreci.com},\\ Gahee Joo\thanks{Research Engineer, Space Technology Research Center, ghjoo@satreci.com}~,~ Seungho Shin\thanks{Research Engineer, Space Technology Research Center, sh\_shin@satreci.com}~,~~ Se Lee\thanks{Research Engineer, Space Technology Research Center, sallyselee@snu.ac.kr}~,~~and Yunhwang Jeong\thanks{Research Engineer, Space Technology Research Center, yhjung@satreci.com}\\%
{\normalsize\it{Satrec Initiative, Daejeon, 34054, Republic of Korea}}\\%
}

\begin{document}

\maketitle

\begin{abstract}

Atmosphere-breathing electric propulsion (ABEP) is a promising technology for long-term orbit maintenance in very-low-Earth orbit. The intake device plays a crucial role in capturing and supplying propellant, and its capture efficiency is a key indicator of drag-compensation feasibility. For experimental evaluation, an EP plasma plume can be used as a particle-flow generator to simulate the VLEO atmosphere in ground facilities. This study numerically investigates the interaction of an EP plasma plume with an intake device to establish guidelines for measuring capture efficiency in conventional vacuum facilities. A hybrid PIC–DSMC method with ion–surface interaction models is employed to simulate the plasma plume incident on the intake. The composition of the captured flow is governed by beam ion energy and species mass: lowering the energy and using lighter atmospheric constituents increase plume divergence and promote neutralization, yielding a neutral-dominated outlet flow. Sputtering of the intake surface becomes non-negligible at high energies but can be mitigated by operating at appropriately low beam energies. The results demonstrate that simultaneous ion and neutral diagnostics are required for reliable capture-efficiency evaluation when using EP plasma plumes in ground facilities.
\end{abstract}

\newpage

\section*{Nomenclature}

\begin{multicols}{2}
\begin{tabular}{ll}
$Kn$ & Knudsen number \\
$\lambda$ & mean free path \\
$L$ & characteristic length \\
$s$ & speed ratio \\
$v$ & velocity \\
$T$ & temperature \\
$k_B$ & Boltzmann constant \\
$m_p$ & particle mass \\
$\dot{N}$ & particle flow rate \\
$n$ & number density  \\
$\eta_c$ & capture efficiency \\
$CR$ & compression ratio \\
$f$ & distribution function \\
$\vec{F}$ & external force \\
$\sigma$ & collision cross-section \\
$\Omega$ & solid angle \\
$\phi$ & plasma potential \\
$e$ & elementary charge \\
$\gamma$ & specific heat ratio \\
$p$ & momentum \\
\end{tabular}

\columnbreak

\begin{tabular}{ll}
\multicolumn{2}{l}{Subscripts} \\
$\infty$ & freestream condition \\
$in$ & inlet condition \\
$out$ & outlet condition \\
$c$ & collision pair \\
$r$ & relative properties \\
$mex$ & momentum-exchange \\
$cex$ & charge-exchange \\
$e$ & electron properties \\
$i$ & ion properties \\
$n$ & neutral atom properties \\
$ref$ & reference properties \\
$\parallel$ & parallel component to the surface  \\
$\perp$ & normal component to the surface  \\ \\

\multicolumn{2}{l}{Superscripts} \\
$*$ & post-collisional properties \\
\end{tabular}
\end{multicols}

\section{Introduction}

Atmosphere-breathing electric propulsion (ABEP) is an emerging space propulsion system for efficient orbit maintenance in very-low-Earth-orbit (VLEO) below 450 km, where enhanced optical resolution and reduced launch cost are obtainable~\cite{Filatyev2023}. Similar to air-breathing engines for aircraft, the ABEP system captures and compresses atmospheric flow in the ram direction using an intake device~\cite{Moon2023vac}. The captured atmospheric gas is then transferred to the ABEP thruster as a propellant. The atmospheric propellant is ionized and accelerated inside the thruster by electromagnetic force, generating thrust for drag compensation maneuvers. Through \textit{in situ} acquisition of propellant, ABEP can facilitate long-term missions in VLEO without depleting onboard propellant~\cite{Schonherr2015,Jun2024rgd}. ABEP is considered to be competitive with conventional electric propulsion (EP) in an altitude range of 160--250 km~\cite{Ko2023}.

The ABEP captures the rarefied VLEO atmosphere, and its degree of rarefaction can be represented with the Knudsen number $Kn$: 
\begin{equation}\label{eqn1}
  Kn = \lambda/L,
\end{equation}
which is a ratio between the mean free path $\lambda$ and characteristic length $L$. The flow regime in VLEO atmosphere is characterized as free molecular flow (FMF), having $Kn$ greater than 10 above 160 km altitude~\cite{Moon2023vac}. VLEO satellites have orbital velocities exceeding 7 km/s, which is more than five times faster than the thermal velocity of atmospheric particles. The speed ratio $s$ is defined as the ratio of the freestream velocity $v_\infty$ to the most probable thermal velocity $v_{mp}$:
\begin{equation}\label{eqn2}
  s=\frac{v_\infty}{v_{mp}}=\frac{v_\infty}{\sqrt{\frac{2k_B T_\infty}{m_p}}}.
\end{equation}
The flow is categorized as a thermal flow for a speed ratio $s$ $<$ 5, where Brownian motion appears significantly in the particle kinetics. On the other hand, the flow becomes hyperthermal when $s$ $>$ 5. The collective motion is predominant, rather than the random motion of the particles in the hyperthermal flow. The freestream to the ABEP intake device is a hyperthermal flow wherein the velocity of the individual gas particles is highly collimated in the ram direction. Intermolecular collision and gas-surface interaction can scatter the particles, reducing the speed ratio $s$. The hyperthermal freestream turns into a thermal flow as the bulk motion diminishes and the random motion intensifies. This process is called thermalization and provides a compression effect for the intake device of ABEP by stagnating the freestream particles~\cite{Moon2023vac}.

Developing an efficient intake device is crucial for realizing a feasible ABEP system for complete drag compensation maneuver~\cite{Moon2024ast,Moon2024rgd}. The capturing process of hyperthermal and rarefied gas flows has been extensively investigated in previous studies using numerical methods, resulting in the development of various optimal intake device designs aimed at achieving high performance~\cite{wu2022progress,Shoda2023,Nishiyama2003,Moon2023vac,Jackson2018,Romano2021,Zheng2021,Zheng2022,Rapisarda2023,Levchenko2020,Jin2024,Fontanarosa2024,Yakunchikov2025,Li2015,Moon2024ast,moon2025,Yi2025,Moon2022}. The intake performance can be measured with compression ratio $CR$ and capture efficiency $\eta_c$,
\begin{equation}
	CR = \frac{n_{out}}{n_{\infty}}\label{eq:CR}.
\end{equation}
\begin{equation}
	\eta_c = \frac{\dot{N}_{out}}{\dot{N}_{in}} = \frac{n_{out}u_{out}A_{out}}{n_{\infty}u_{\infty}A_{in}}\label{eq:eta_c},
\end{equation}
The compression ratio $CR$ represents the ratio of the number density of the captured propellant to that of the freestream atmosphere. The capture efficiency $\eta_c$ represents the flow rate ratio of the captured gas entering the thruster to the total atmospheric freestream incident on the intake device's inlet area. $\eta_c$ is a critical parameter for assessing drag-compensation feasibility, as it quantifies the capability of supplying propellant to the ABEP thruster.

Beyond numerical characterization, ground tests have been conducted to validate both the functionality of intake devices and the feasibility of the ABEP concept~\cite{Andreussi2022}. Electric propulsion (EP) systems can be employed as particle flow generators (PFGs) to simulate rarefied, hyperthermal flow with a consistent incident flux to intake devices~\cite{Hruby2022,Andreussi2022,Andreussi2022CEAS,Prochnow2022}. Unlike the predominantly neutral VLEO atmosphere, the charged particles in EP plasma plumes can interact with the surface of intake devices through mechanisms such as electric acceleration, sheath formation, and neutralization~\cite{lieberman2005}. Moreover, direct impingement of high-speed ions might induce sputtering~\cite{Choi2017}. Accounting for these altered surface interactions is essential when evaluating intake performance in ground tests employing EP plasma plumes.


The ground tests can be classified into end-to-end testing and stand-alone intake device testing. End-to-end testing employs an integrated intake device–thruster assembly, and several campaigns have demonstrated ABEP operation by generating thrust from the captured gas flow~\cite{Tagawa2013,Andreussi2022CEAS,Hruby2022}. However, characterizing intake performance remains challenging in end-to-end testing, as the flow properties at the outlet of the intake device cannot be measured separately. Recently, stand-alone intake device testing apparatus have been proposed to investigate intake performance~\cite{cushen2024,Parodi2024}. These apparatus enable direct measurement of the captured flow rate for evaluating $\eta_c$, but require specialized facilities with segmented auxiliary reservoirs and pumps in addition to conventional vacuum chambers. If intake performance could be measured in conventional vacuum chambers, stand-alone intake device testing would be more accessible using existing diagnostics; however, prior understanding of the captured flow is required to select appropriate techniques.

This study numerically investigates the EP plasma plume flow into an intake device to establish guidelines for evaluating $\eta_c$ in conventional vacuum chambers. The exhaust plasma plume of a gridded ion thruster is simulated under facility background pressure and directed into the intake device to assess outlet flow characteristics. The analysis considers variations in beam ion energy and species mass. A hybrid Particle-in-Cell/Direct Simulation Monte Carlo (PIC–DSMC) method is employed, incorporating a polytropic electron fluid model and physical models for intermolecular collisions, gas-surface interaction, and ion-surface interactions.

\section{Physical Models and Numerical Methodology}

The rarefied gas flow and collisional plasma plume flow can be described by the Boltzmann equation (BE):
\begin{equation}
\frac{\partial}{\partial t}(nf) + \vec{v}\cdot \frac{\partial}{\partial \vec{r}}(nf)+\frac{\vec{F}}{m_p}\cdot\frac{\partial}{\partial \vec{v}}(nf)=\int_{-\infty}^{\infty}\int_{0}^{4\pi}n^2[f^*f^*_c-ff_c]v_r\sigma  d\Omega  d\vec{v}_1,
\label{eq:boltzmann}
\end{equation}
where the BE governs the time evolution of the particle distribution function $f$ in phase space, spanned by position $\vec{r}$ and velocity $\vec{v}$. The right-hand side represents the binary collision operator, while the left-hand side accounts for particle advection under the influence of external forces $\vec{F}$, including the self-consistent electric field arising from the collective motion of charged particles.

Momentum exchange (MEX) collisions between neutral atoms are modeled using the variable hard sphere (VHS) model, which assumes isotropic scattering with a total cross section (TCS) dependent on the relative velocity $v_r$ of the colliding pair~\cite{bird1994molecular}. Following MEX collisions are simulated with VHS model:
	\begin{equation}
		Xe(p_1)+Xe(p_2)\rightarrow Xe(p_1')+Xe(p_2'),
		\label{eq:mex1}
	\end{equation}
	\begin{equation}
		O(p_1)+O(p_2)\rightarrow O(p_1')+O(p_2').
		\label{eq:mex2}
	\end{equation}
In contrast, ion–atom collisions are characterized as long-range charge–induced dipole interactions, where small-angle scattering dominates. Accurate simulation of ion–atom interactions requires not only TCS but also detailed modeling of the differential cross section (DCS). This study employs the generalized collision cross-section Model (GCCM) developed for monatomic ion–atom interactions~\cite{Moon2023psst}. GCCM evaluates both TCS and DCS as functions of atomic number and collision energy, enabling the accurate determination of deflection angles and post-collisional momenta for ions and atoms, respectively. Following ion-atom MEX collisions are modeled with GCCM:
	\begin{equation}
		Xe^+(p_1)+Xe(p_2)\rightarrow Xe^+(p_1')+Xe(p_2'),
		\label{eq:mex4}
	\end{equation}
	\begin{equation}
		O^+(p_1)+O(p_2)\rightarrow O^+(p_1')+O(p_2').
		\label{eq:mex5}
	\end{equation}
In addition to the MEX process, ion-atom collisions can involve charge exchange (CEX):  
	\begin{equation}
		Xe^+(p_1)+Xe(p_2)\rightarrow Xe(p_1)+Xe^+(p_2),
		\label{eq:cex1}
	\end{equation}
	\begin{equation}
		O^+(p_1)+O(p_2)\rightarrow O(p_1)+O^+(p_2).
		\label{eq:cex2}
	\end{equation}
For the resonant CEX of Eqs.~(\ref{eq:cex1}) and (\ref{eq:cex2}), the TCS of CEX is assumed to be equivalent to that of the MEX estimated by GCCM~\cite{Boyd2002,Tumuklu2018,Miller2002,Andrews2020,Stephani2014,Rapp1962}.

Gas-surface interaction (GSI) of neutral atoms is modeled using the Maxwell model, which describes the kinetic energy of reflected particles through the energy accommodation coefficient $\alpha_\epsilon$: 
\begin{equation} 
    \alpha_\epsilon = \frac{T_{k,pre} - T_{k,post}}{{T_{k,pre}}-T_{s}}, 
    \label{eq:alpha} 
\end{equation} 
where $T_{k,pre}$ and $T_{k,post}$ represent the kinetic energy of a particle before and after the GSI, respectively, while $T_{s}$ denotes the surface temperature where the particle accommodates. A specular reflection occurs when $\alpha_\epsilon = 0$, whereas a fully diffuse reflection takes place when $\alpha_\epsilon = 1$. On the other hand, ion–surface interaction (ISI) is modeled for scattering with sheath acceleration, neutralization, and sputtering. In the vicinity of a surface, a plasma sheath develops due to the mobility difference between ions and electrons~\cite{lieberman2005}. The sheath potential $\phi_{sh}$ between plasma and a floating surface can be derived from the impinging ion and electron current balance:
\begin{equation} 
    J_e=J_{i,\perp}.
    \label{eq:current_bal} 
\end{equation} 
Here, the ion current density normal to the surface is denoted by $J_{i,\perp}$, while the electron current density is represented by $J_e$. Considering the half-Maxwellian flux of electrons at $T_e$, the sheath potential $\phi_{sh}$ can be expressed as a function of $J_{i,\perp}$:
\begin{equation}
\phi_{sh} = \frac{k_B T_e}{e}\left[\ln\left(\frac{4J_{i,\perp}}{e n_e}\right) - \frac{1}{2}\ln\left(\frac{8k_B T_e}{\pi m_e}\right)\right].
\label{eq:phi_sh}
\end{equation}
Using the calculated $\phi_{sh}$, the terminal ion velocity components $v'_{i,\perp}$ and $v'_{i,\parallel}$ incident on the floating surface are corrected as:
\begin{equation}
v_{i,\parallel}' = v_{i,\parallel},
\label{eq:v_para}
\end{equation}
\begin{equation}
v_{i,\perp}' = \sqrt{v_{i,\perp}^2 + \frac{2e\phi_{sh}}{m_i}}.
\label{eq:v_perp}
\end{equation}
After passing through the sheath, ions impinging on the surface may undergo scattering, neutralization, or sputtering. Scattering is modeled using the Maxwell model, while neutralization is applied with a probability $P_{neut}$, producing reflected neutral atoms. Sputtering of surface atoms depends on the incident ion energy $E$ and grazing angle $\theta$. The total sputtering yield $Y(E,\theta)$ is expressed as the product of the energy-dependent yield $Y(E,0)$ and the angular-dependent yield $Y'(\theta)$:
\begin{equation}
Y(E,\theta) = Y(E,0)\cdot Y'(\theta).
\label{eq:Y_tot}
\end{equation}
Semi-empirical formulae suggested in Yim's paper are employed to compute the sputtering yield, which are Eckstein's energy dependence model and reformulated Wei's angular dependence model~\cite{Yim2017,Eckstein2003,Wei2008}: 
\begin{equation}
Y(E,0)=Qs_n\frac{\left(\frac{E}{E_{th}}-1\right)^\mu}{\frac{\lambda}{w}+\left(\frac{E}{E_{th}}-1\right)^\mu},
\label{eq:eckstein}
\end{equation}
\begin{equation}
Y'(\theta) = \frac{1}{1+(\beta/\alpha)^2\tan^2\theta}\exp\left(\frac{1}{2}\left(\frac{a}{\alpha}\right)^2\left[1-\frac{1}{1+(\beta/\alpha)^2\tan^2\theta}\right]\right).
\label{eq:v_perp}
\end{equation}
Here, ($Q$, $\lambda$, $\mu$, $E_{th}$) and ($\beta/\alpha$, $a/\alpha$) are fitting parameter sets for the respective models. The reduced nuclear stopping power $s_n$ is formulated from the Krypton–Carbon (KC) potential, and $w$ is a parameter dependent on reduced energy~\cite{Yim2017}. The number of sputtered atoms is determined from $Y(E,\theta)$, and their energies $E_{spt}$ are sampled according to the Sigmund–Thompson distribution~\cite{Sigmund1981}:
\begin{equation}
f(E_{spt})\propto \frac{E_{spt}}{(E_{spt}+U_B)^{3-2l}},
\label{eq:v_perp}
\end{equation}
where $U_B$ is the surface binding energy and $l$ is the interatomic potential exponent~\cite{Choi2017}. The angular distribution of sputtered atoms leaving the surface is assumed to follow a cosine law~\cite{Choi2017}.

A hybrid kinetic simulation method is employed to obtain a stochastic solution of the Boltzmann equation (BE) with the physical models, combining two complementary approaches: Particle-in-Cell (PIC) and Direct Simulation Monte Carlo (DSMC). The hybrid PIC–DSMC method treats ions and atoms as discrete particles, whereas electrons are modeled as a fluid. The PIC and DSMC methods work in tandem to compute electrostatic accelerations and collisions that govern particle motion over time. The DSMC method provides the primary framework through particle sorting, advection, and collision procedures, while the PIC method supplies the electrostatic acceleration of charged particles via charge density assignment, field updating, and field interpolation~\cite{Moon2023psst}. The PIC method calculates the plasma potential from the spatial charge density and a polytropic cooling model of the electron fluid, whereby the electron momentum equation is reduced to~\cite{Tekinalp2019,Boyd2002,Merino2015,Moon2023psst,Moon2022plume}:
  	\begin{equation}
		\begin{aligned}
			\phi = 
			\begin{cases}
				\displaystyle \phi_{ref} + \frac{k_B T_{e,ref}}{e}\ln\left(\frac{n_e}{n_{e,ref}}\right) \quad & \text{for}\quad\gamma = 1,\\[2ex]
				\displaystyle \phi_{ref} + \frac{k_B T_{e,ref}}{e}\frac{\gamma}{\gamma-1}\left[\left(\frac{n_e}{n_{e,ref}}\right)^{\gamma-1}-1\right] \quad & \text{for}\quad\gamma > 1 .
			\end{cases}
		\end{aligned}
		\label{eq:poly_pot}
	\end{equation}
This study adopts a specific heat ratio of $\gamma = 1.3$ for the plasma plume flow, and charge quasi-neutrality is assumed as $n_e \approx n_i$~\cite{Merino2020,Moon2023psst}. The electric field is obtained by finite differencing of the potential and interpolated to individual particles to assign the corresponding electrostatic acceleration. The hybrid PIC–DSMC code employed in this study is adopted from Moon \textit{et al.}, developed on the basis of the SPARTA DSMC framework\cite{Moon2023psst,Plimpton2019}.

\section{Simulation Settings}

\begin{figure}[htb!]
  \centering
  \includegraphics[width = \textwidth]{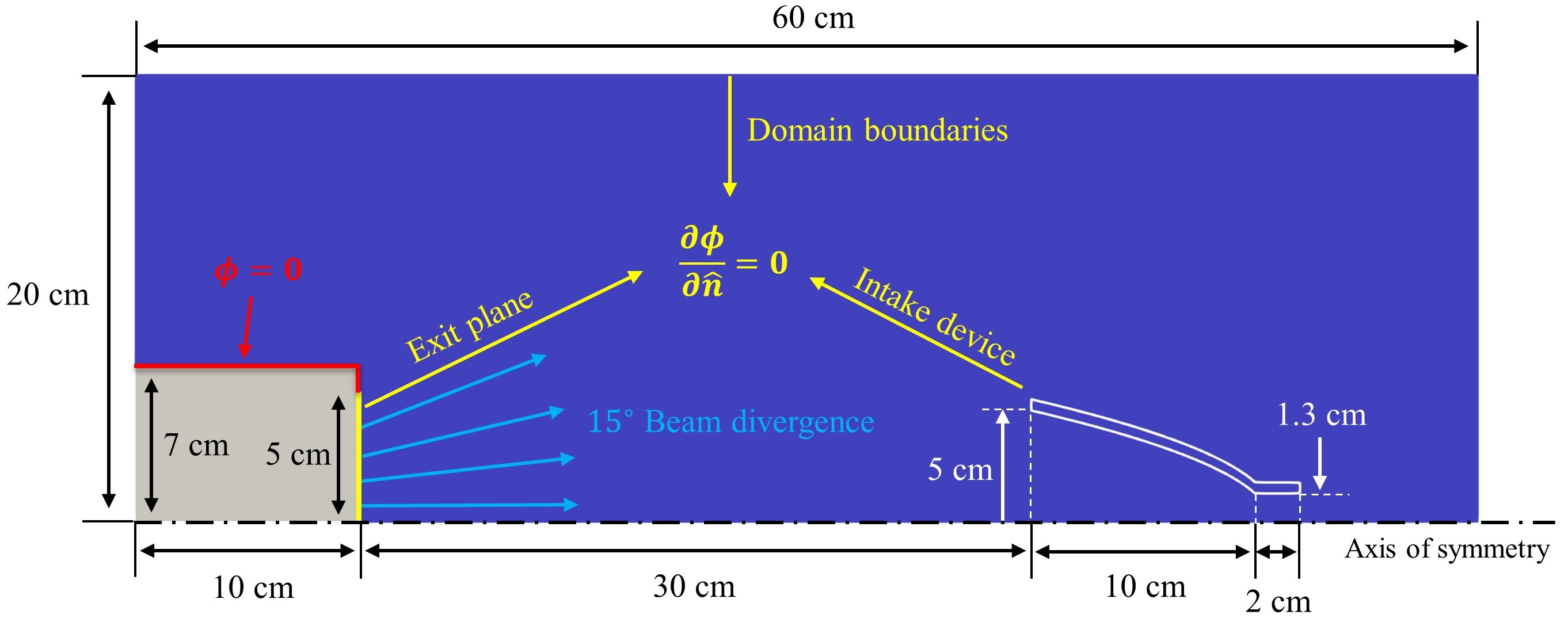}
  \caption{Simulation domain}
  \label{fig:domain}
\end{figure}

The plasma plume of a gridded ion thruster is simulated in an axisymmetric domain with 60 cm in length and 20 cm in radius, as shown in Fig.~\ref{fig:domain}. The thruster exit plane has a diameter of 10 cm. Ions are exhausted from the exit plane with a divergence angle of $15^{\circ}$, and the ion number density profile is prescribed as a bivariate Gaussian distribution with respect to the radial position. Multiply charged ions are neglected. Neutral atoms are emitted with a half-Maxwellian velocity distribution at $T_n = 500$ K. Four simulation cases are defined according to the exhaust conditions, varying ion energy and propellant species. The Xe 1000 eV case represents the conventional operating condition of ion thrusters with xenon propellant. The Xe 250 eV case corresponds to a halved ion velocity $v_i$, while the Xe 40 eV case yields $v_i$ consistent with the VLEO freestream velocity. To investigate the effect of propellant substitution with a lighter atmospheric constituent, the O 5 eV case is defined using atomic oxygen propellant, with $v_i$ also consistent with the VLEO freestream velocity. The cases and exhaust conditions are summarized in Table.~\ref{tab:case_table}. All cases are configured with a constant flow rate of 3.3 sccm and an ionization degree of 90\%. A facility background pressure of $5\times10^{-6}$ torr is imposed in the simulations, with neutral particles at 300 K injected through the domain boundaries.
\begin{table}[h]
  \centering
  \caption{Exhaust conditions of ion thruster}
  \label{tab:case_table}
  \begin{tabular}{llcccc}
    \toprule
    \textbf{Properties} & \textbf{Xe 1000 eV} & \textbf{Xe 250 eV} & \textbf{Xe 40 eV} & \textbf{O 5 eV} \\
    \midrule
    Species & \multicolumn{3}{c}{$Xe$/$Xe^+$} & $O$/$O^+$  \\
    $n_{n}$ [m$^{-3}$]    &          & $1.00\times10^{17}$ &        & $3.48\times10^{16}$ \\
    $v_{n}$ [m/s]       &          & 284        &        & 815  \\
    $T_{n}$ [K]   &  \multicolumn{4}{c}{500}      \\
    $n_{i}$ [m$^{-3}$] & $4.55\times10^{15}$ & $9.10\times10^{15}$ & $2.28\times10^{16}$ & $2.28\times10^{16}$  \\
    $v_{i}$ [m/s] & $3.9\times10^{4}$   & $1.95\times10^{4}$  & \textbf{$7.8\times10^{3}$} & $7.8\times10^{3}$ \\
    $T_{i}$ [eV]  & \multicolumn{4}{c}{1}      \\
    \bottomrule
  \end{tabular}
\end{table}
At 30 cm downstream from the thruster exit plane, a parabolic intake device is placed, with an inlet radius of 5 cm, an outlet radius of 1.3 cm, and a total length of 12 cm. The intake device is assumed to be composed of pyrolytic graphite, applying GSI and ISI properties specified in Table~\ref{tab:surf}. For the O 5 eV case, the ion energy is lower than the surface binding energy; therefore, sputtering is neglected for $O^+$ ion impingement. The GSI and neutralization of ISI are also considered to the thruster's surface. 

\begin{table}[h]
    \centering
    \begin{threeparttable}
    \caption{GSI and ISI properties}
                \label{tab:surf}
    \begin{tabular}{cccc}
        \toprule
        \textbf{Interaction} & \textbf{Parameter} & \textbf{Value} & \textbf{ref.} \\
        \midrule
        \multirow{2}{*}{GSI} & $\alpha_\epsilon$ & $1.0$ & $^{\star}$  \\ 
        & $T_s$ & 500 K & $^\star$ \\ \hline
        \multirow{9}{*}{ISI} & $P_{neut}$  & $1.0$ & $^{\star}$ \\
        & $Q$         & $4.0$ & \cite{Yim2017} \\
        & $\lambda$   & $0.8$ & \cite{Yim2017} \\
        & $\mu$       & $1.8$ & \cite{Yim2017} \\
        & $E_{th}$    & $21$ & \cite{Yim2017} \\
        & $\beta/\alpha$ & 0.88  & \cite{Yim2017} \\
        & $a/\alpha$ & 2.20 & \cite{Yim2017} \\
        & $U_B$    & $7.4$ eV & \cite{Choi2017} \\
        & $l$         & $1/3$ & \cite{Choi2017} \\
        \bottomrule
    \end{tabular}
    \begin{tablenotes}
        \item[$\star$]assumed properties
    \end{tablenotes}
    \end{threeparttable}
\end{table}

The computational domain is discretized using an adaptively refined five-level hierarchical structured grid. For the PIC calculations, the thruster exit plane and the intake device are treated with Neumann boundary conditions, imposing a zero potential gradient, while the remaining thruster surfaces are grounded. The timestep is set to $5\times10^{-8}$ s to resolve ion plasma oscillation periods throughout the domain. Disparate weighting factors of 1.0 and 0.01 are applied to neutral atoms and ions, respectively. In addition, radial cell weighting is employed to ensure an even distribution of simulation particles within the axisymmetric domain. Each case is iterated up to 800,000 steps, sampling macroscopic flow properties after the flow reaches steady state.

\section{Results}

\begin{figure}[htb!]
	\centering
	\subfloat[Xe 1000 eV.]{\includegraphics[width=8.2cm]{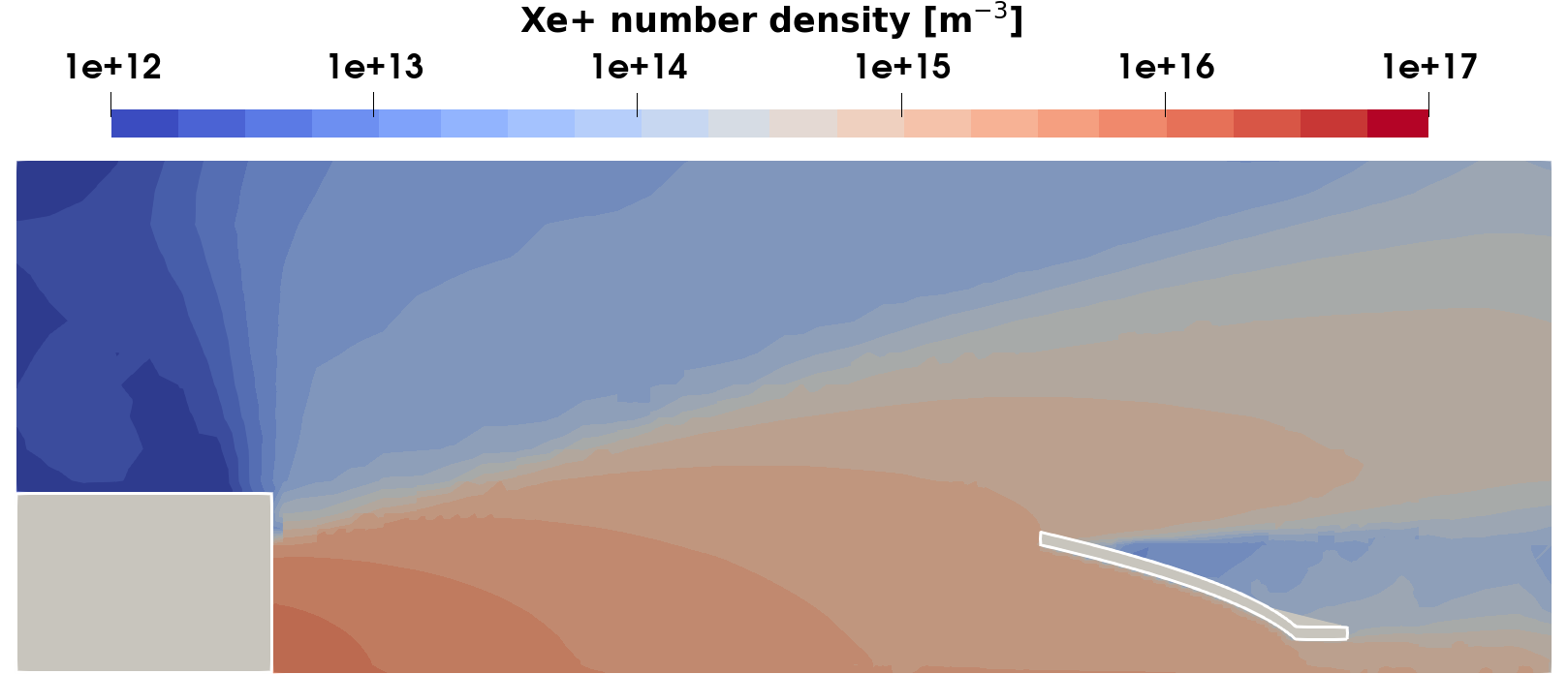}\label{fig:CRAID_concept}}
	\hfill
	\subfloat[Xe 250 eV.]{\includegraphics[width=8.2cm]{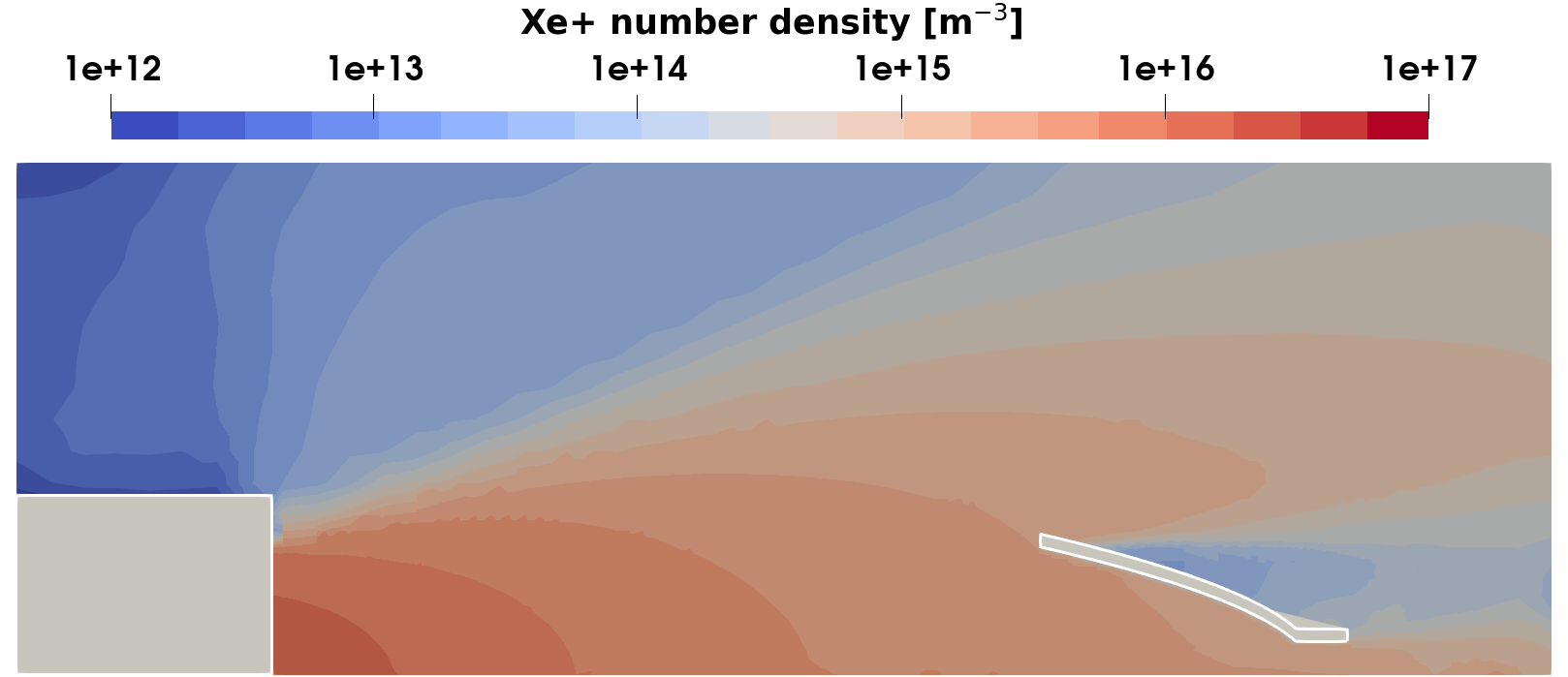}\label{fig:CRAID_concept}}
	\vfill
    	\subfloat[Xe 40 eV.]{\includegraphics[width=8.2cm]{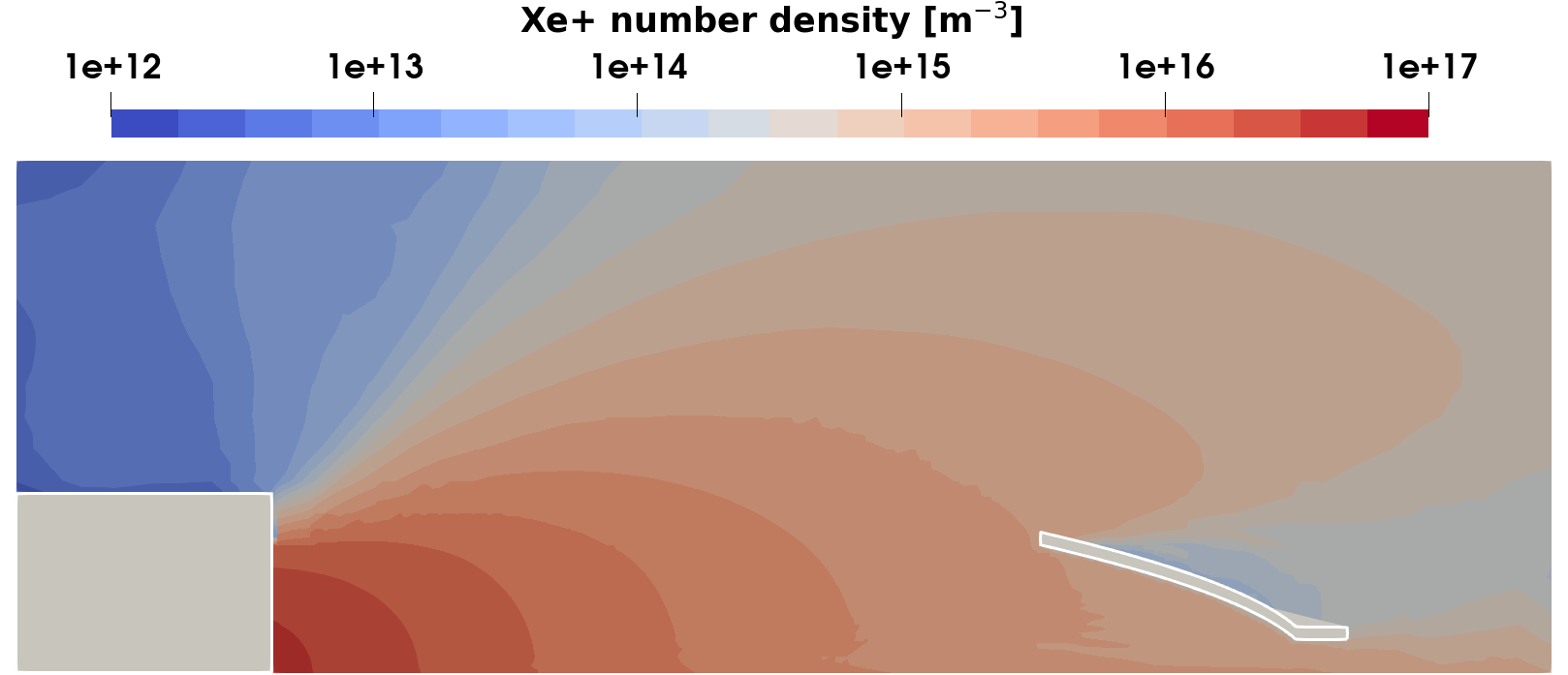}\label{fig:CRAID_concept}}
	\hfill
    	\subfloat[O 5 eV.]{\includegraphics[width=8.2cm]{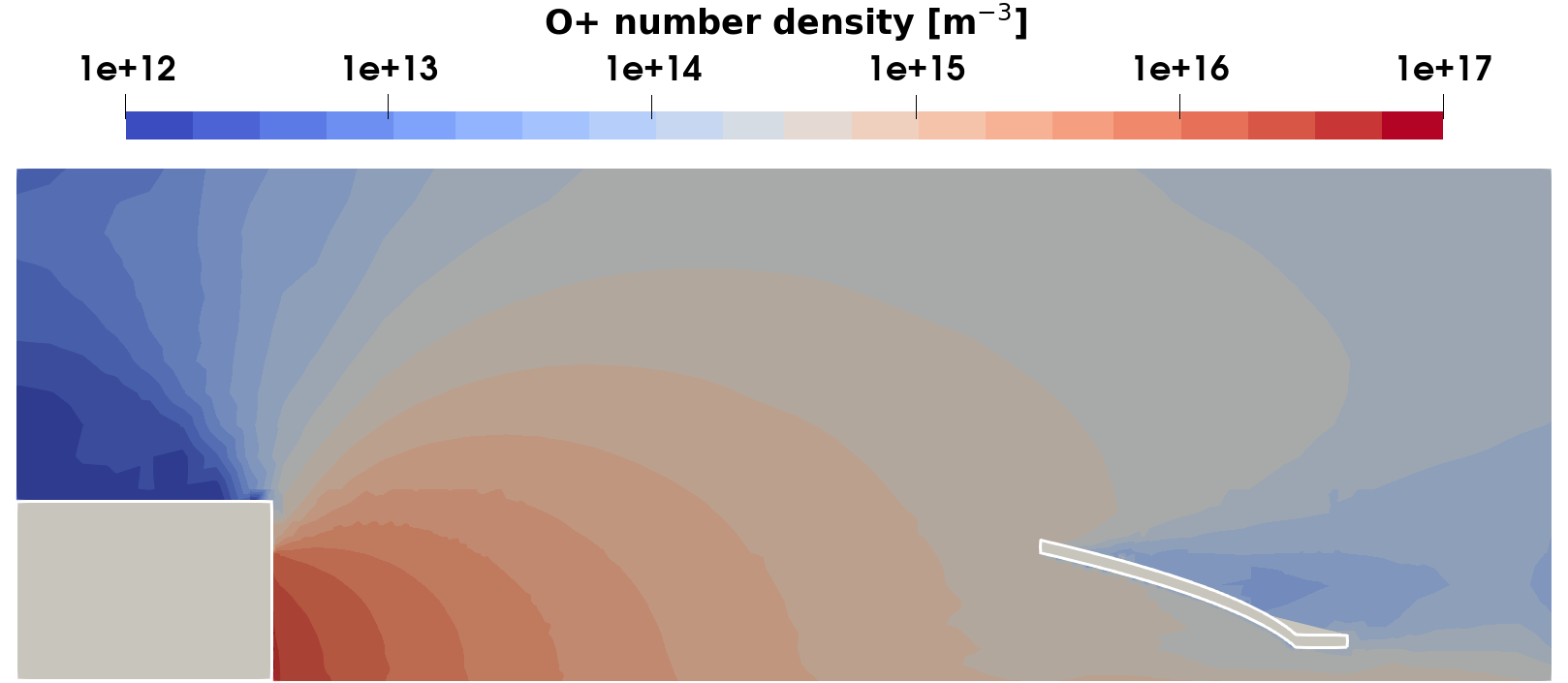}\label{fig:CRAID_concept}}
	
	\caption{Ion number density distribution.}
	\label{fig:ion}
\end{figure}

Fig.~\ref{fig:ion} shows contour diagrams of the ion number density distribution. The ion beam expands downstream and enters the intake device. In Figs.~\ref{fig:ion}(a)–(c), the plume envelope exhibits greater divergence as the exhaust ion energy decreases, since diffusive transport and transverse electric acceleration become more significant due to the higher charge density in the thruster near-field. Owing to the lighter mass of O ions compared with Xe ions, the O 5 eV case in Fig.~\ref{fig:ion}(d) produces a more diffusive plume than the Xe 40 eV case in Fig.~\ref{fig:ion}(c), even though both cases share the same exhaust ion number density and velocity. This implies that reducing ion acceleration and substituting the propellant with atmospheric constituents can improve similarity to VLEO freestream composition; however, the resulting higher divergence may hinder the simulation of hyperthermal flows characterized by highly aligned particle trajectories.

\begin{figure}[htb!]
	\centering
	\subfloat[Xe 1000 eV.]{\includegraphics[width=8.2cm]{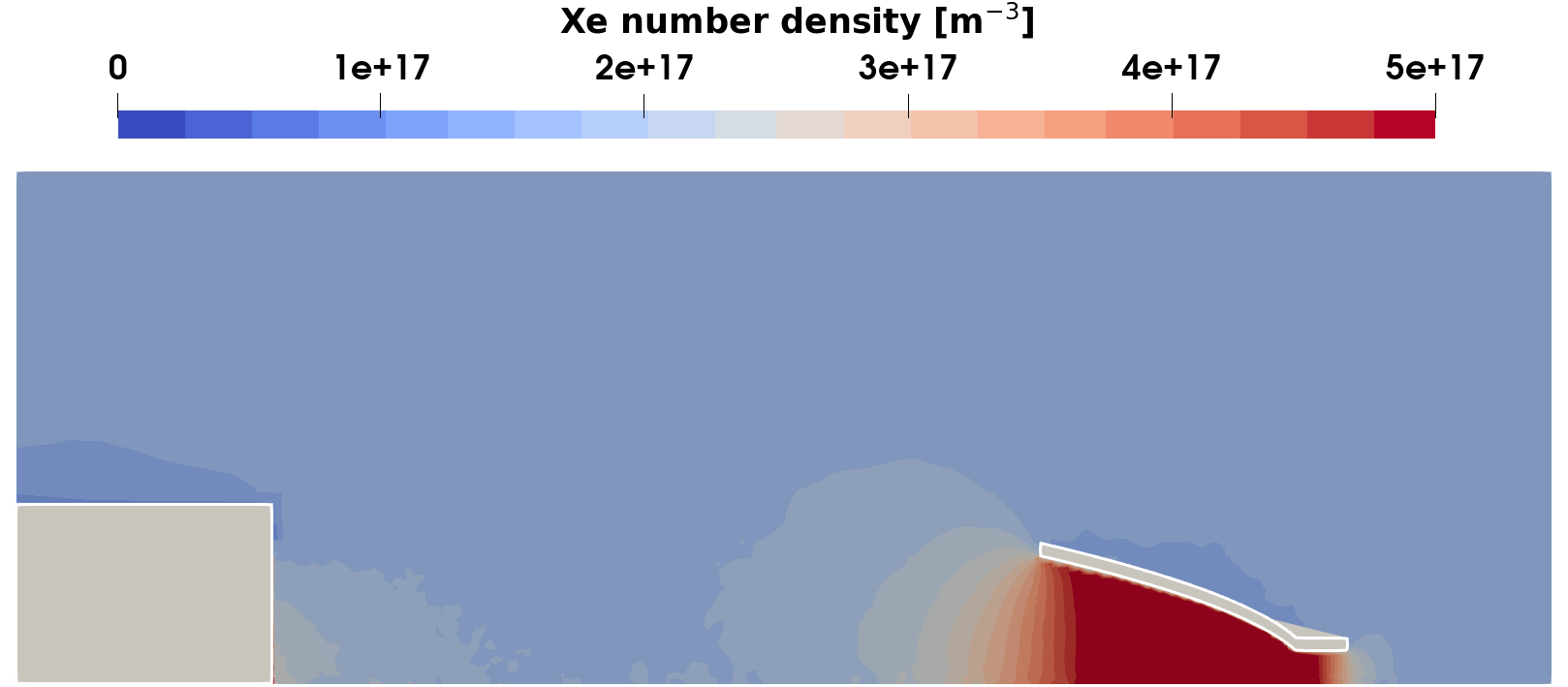}\label{fig:CRAID_concept}}
	\hfill
	\subfloat[Xe 250 eV.]{\includegraphics[width=8.2cm]{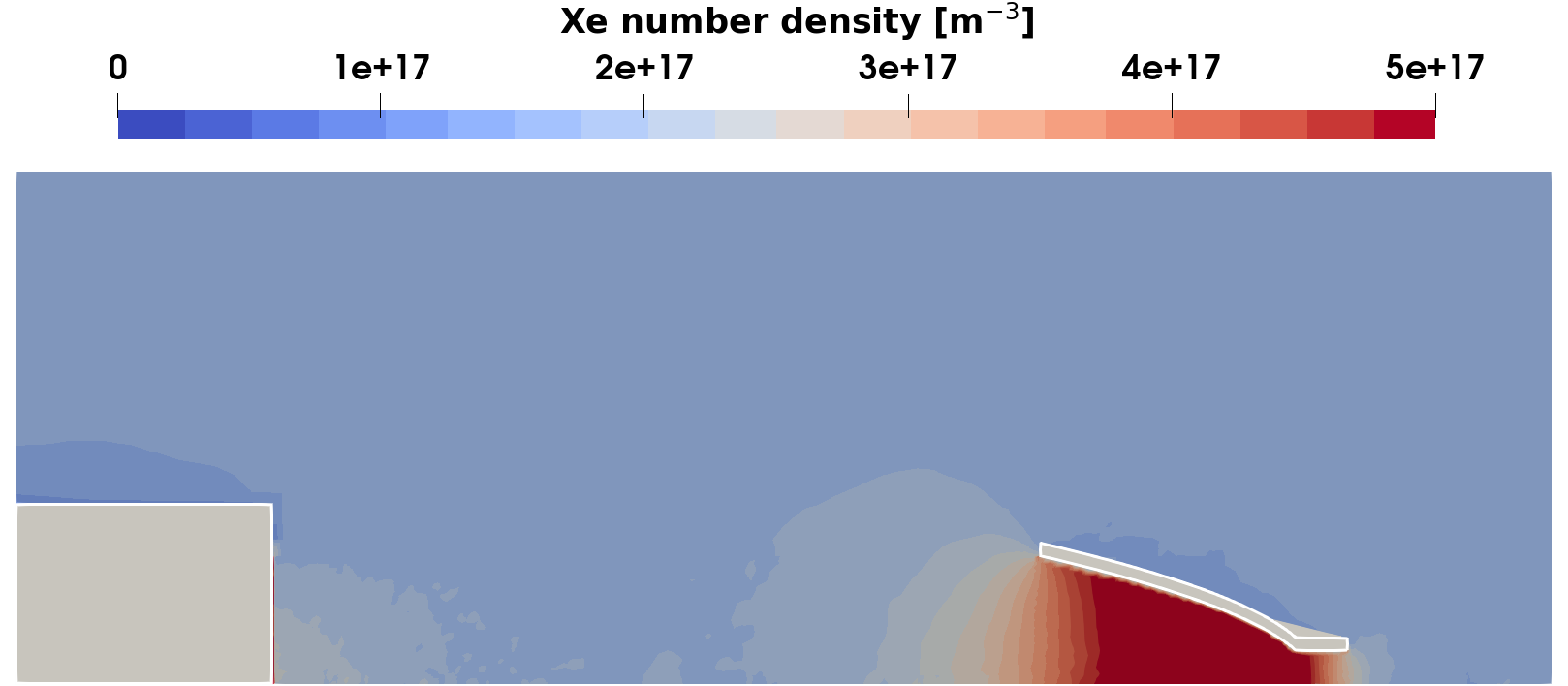}\label{fig:CRAID_concept}}
	\vfill
    	\subfloat[Xe 40 eV.]{\includegraphics[width=8.2cm]{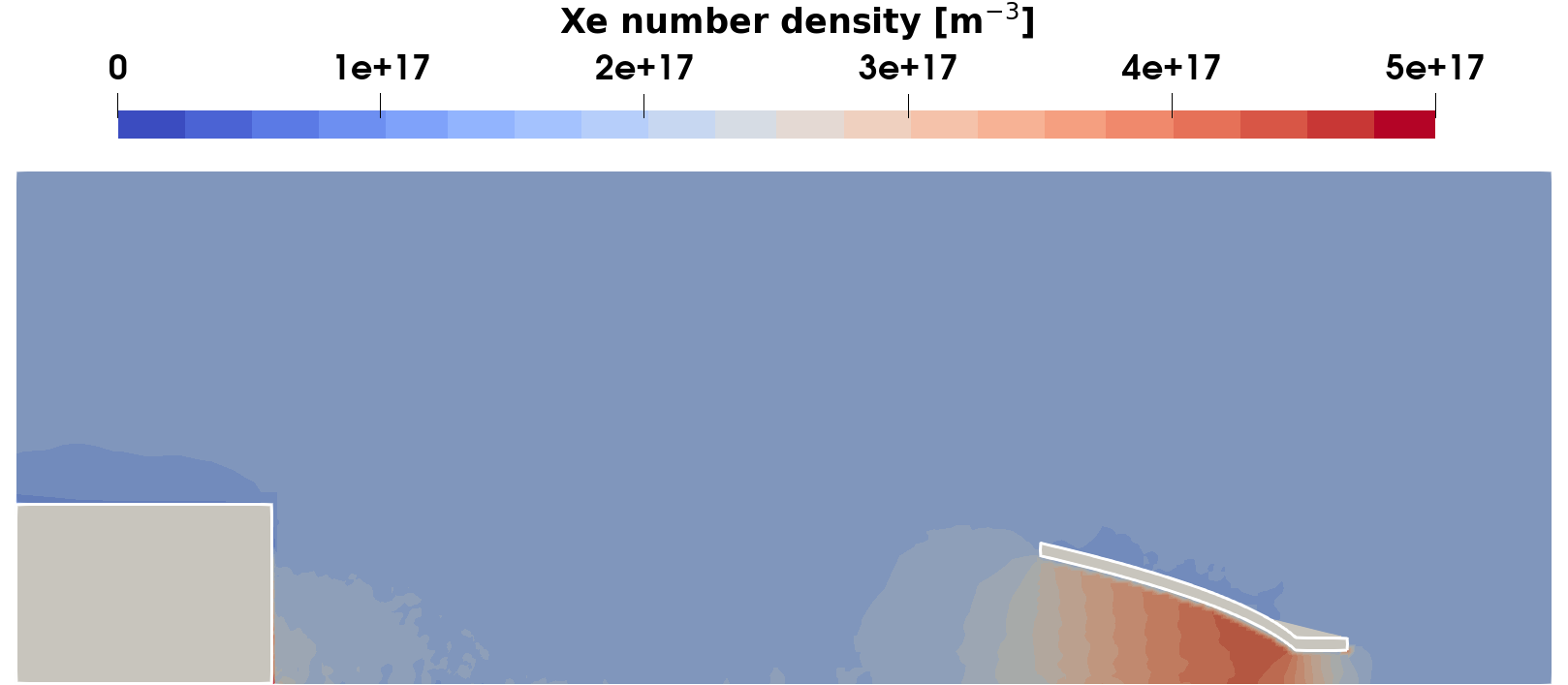}\label{fig:CRAID_concept}}
	\hfill
    	\subfloat[O 5 eV.]{\includegraphics[width=8.2cm]{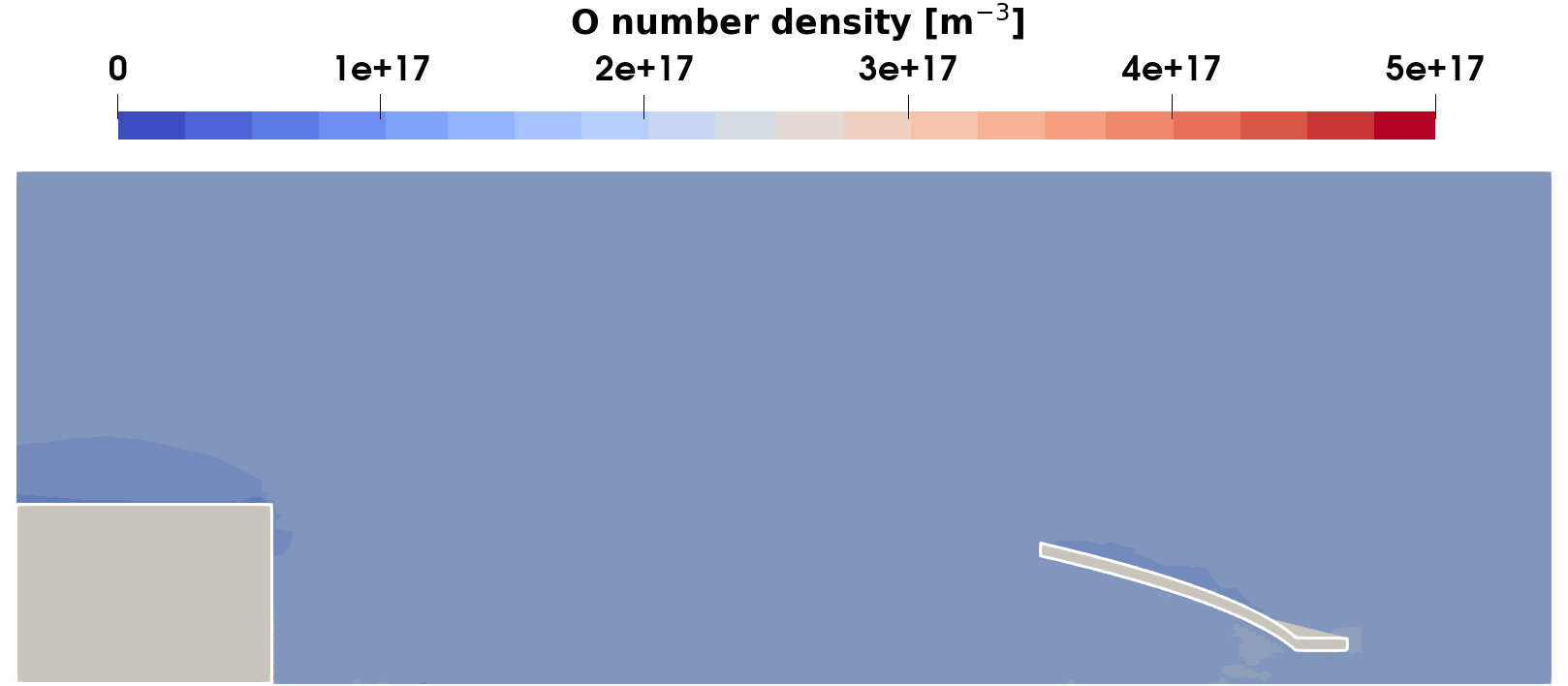}\label{fig:CRAID_concept}}
	
	\caption{Neutral atom number density distribution.}
	\label{fig:neutral}
\end{figure}

As ions impinge on the internal surface of the intake device, they undergo neutralization and form compressed gas inside, as shown in the neutral atom density distribution in Fig.~\ref{fig:neutral}. Since higher exhaust ion energy produces a more collimated beam, a larger ion flux enters the intake device, resulting in a higher concentration of neutral atoms. For the Xe 1000 eV case, the inlet ion flux is $3.96\times10^{19}$ m$^{-2}$s$^{-1}$. This decreases to $3.37\times10^{19}$ m$^{-2}$s$^{-1}$ and $1.65\times10^{19}$ m$^{-2}$s$^{-1}$ for the Xe 250 eV and Xe 40 eV cases, respectively. The O 5 eV case exhibits the lowest inlet ion flux of $7.56\times10^{18}$ m$^{-2}$s$^{-1}$, which is 5.2 times lower than that of the Xe 1000 eV case. Accordingly, the neutral atom concentration inside the intake device is highest for the Xe 1000 eV case and lowest for the O 5 eV case, where the density increase remains less than 10\% compared to the background pressure. Since a large fraction of ions neutralize within the intake device, the neutral atom flux escaping the outlet increases to 4.9–13.4 times the incoming neutral flux at the inlet. In contrast, the ion flux at the outlet decreases to 48.3–62.4\% of the inlet value. The ion and neutral fluxes at the inlet and outlet of the intake device are summarized in Table~\ref{tab:capture_eff}.

\begin{figure}[htb!]
	\centering
	\subfloat[Xe 1000 eV.]{\includegraphics[width=8.2cm]{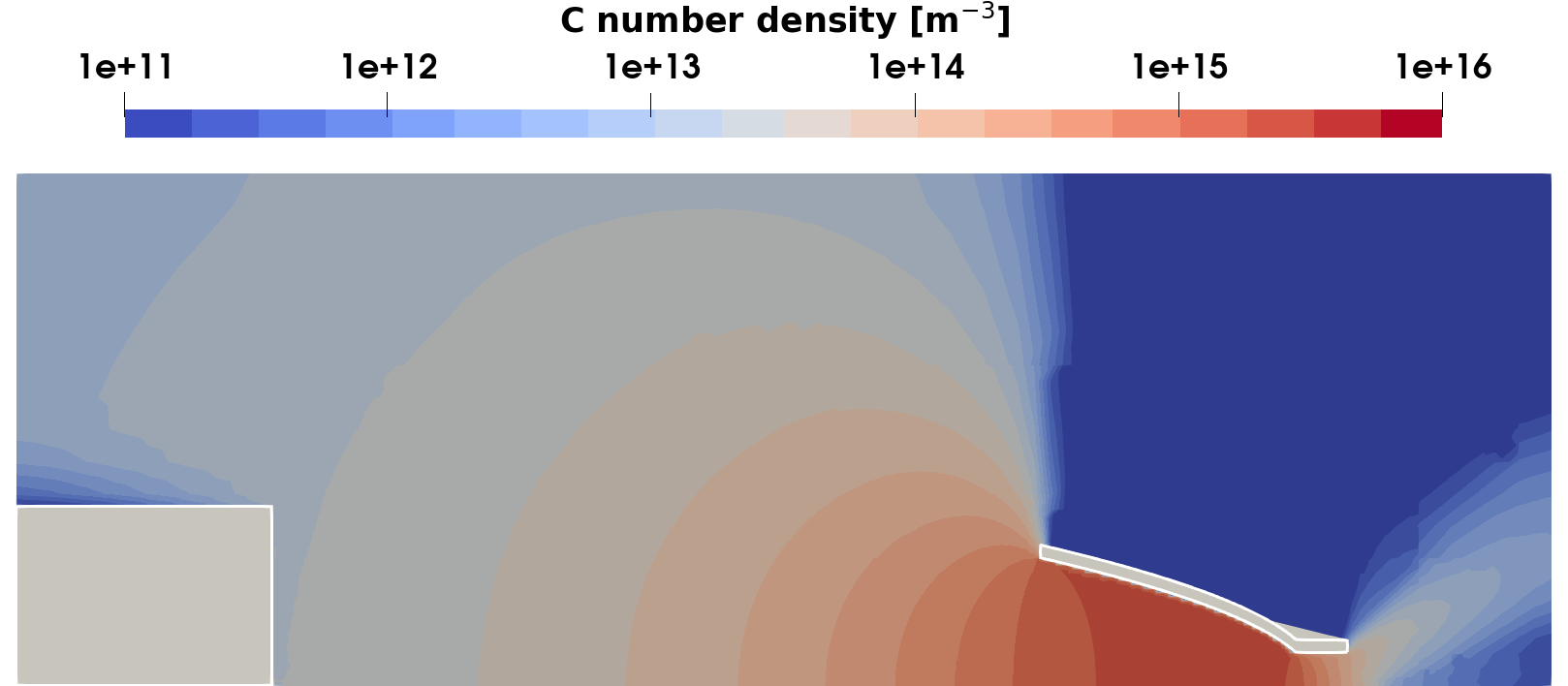}\label{fig:CRAID_concept}}
	\hfill
	\subfloat[Xe 250 eV.]{\includegraphics[width=8.2cm]{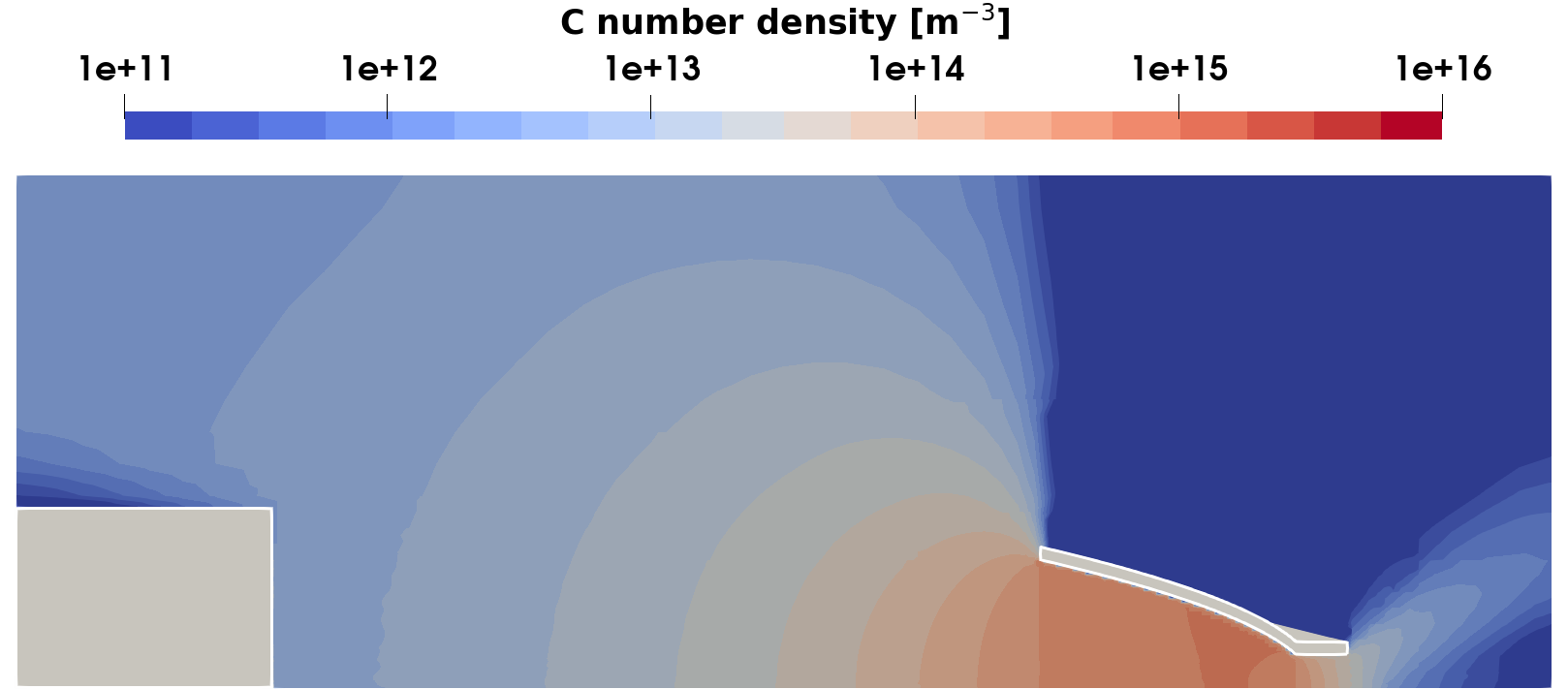}\label{fig:CRAID_concept}}
	\vfill
    	\subfloat[Xe 40 eV.]{\includegraphics[width=8.2cm]{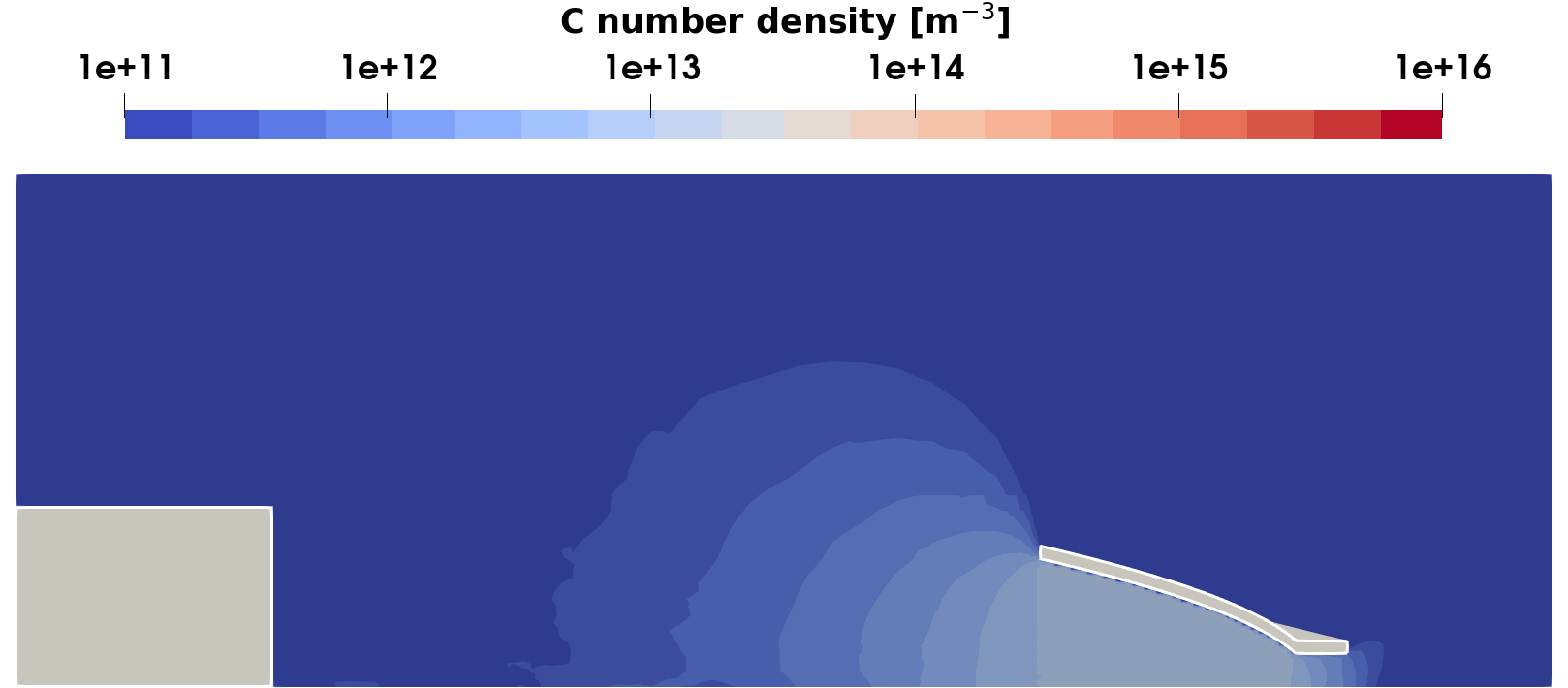}\label{fig:CRAID_concept}}
	\caption{Sputtered C atom number density distribution.}
	\label{fig:sputtered}
\end{figure}

Ion impingement on the intake device surface produces not only neutralized atoms but also sputtered atoms, depending on the incident energy. Fig.~\ref{fig:sputtered} shows the number density distribution of sputtered carbon atoms. For xenon projectiles in the 40–1000 eV energy range, the sputtering yield of pyrolytic graphite increases with ion energy~\cite{Yim2017}. Accordingly, the Xe 1000 eV case in Fig.~\ref{fig:sputtered}(a) exhibits the most significant sputtering of C atoms, whereas the Xe 40 eV case shows a C number density less than 0.5\% of that in the former case. The sputtered C atoms also pass through the outlet of the intake device, with fluxes of $2.24\times10^{18}$, $5.12\times10^{17}$, and $8.14\times10^{15}$ m$^{-2}$s$^{-1}$ for each case, as summarized in Table~\ref{tab:capture_eff}. The outlet flux of sputtered atoms may distort measurements of the captured flow rate through the intake device, thereby affecting the evaluation of capture efficiency $\eta_c$.

\begin{figure}[htb!]
  \centering
  \includegraphics[width = 12cm]{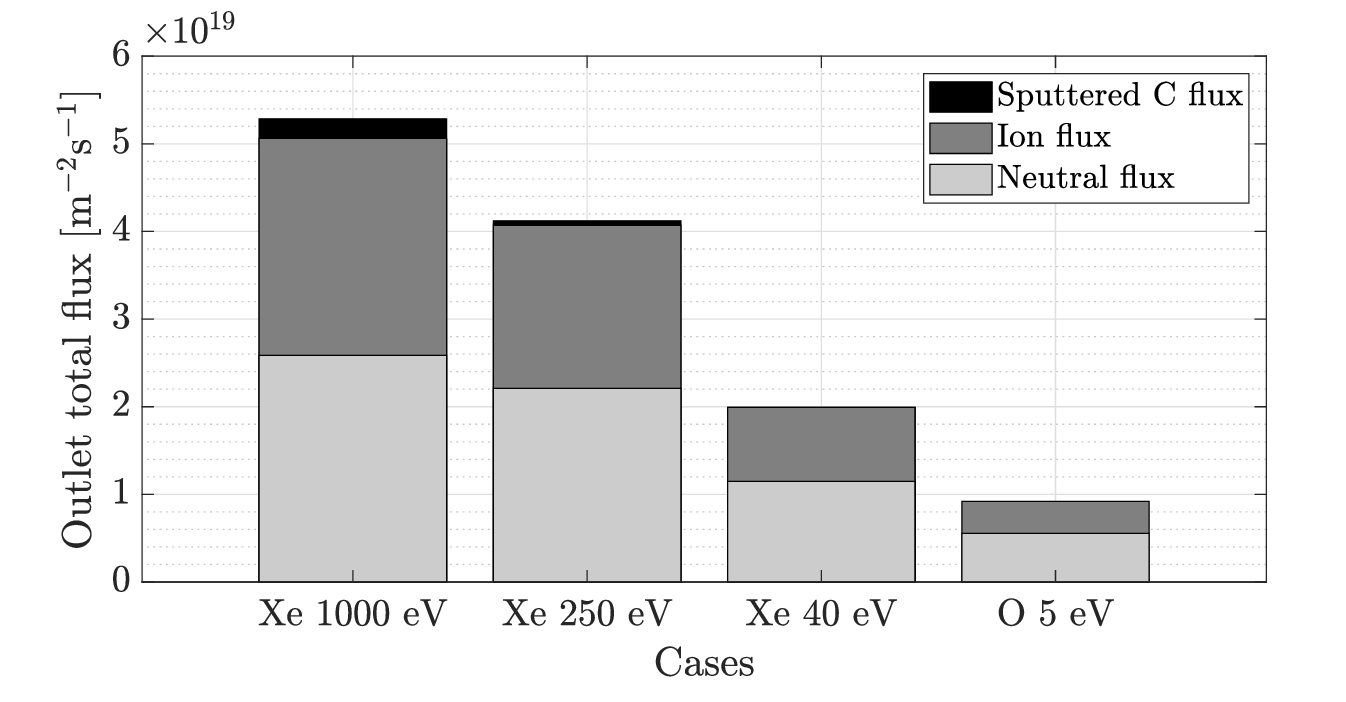}
  \caption{Outlet total flux and its composition}
  \label{fig:flux}
\end{figure}

\begin{figure}[htb!]
  \centering
  \includegraphics[width = 12cm]{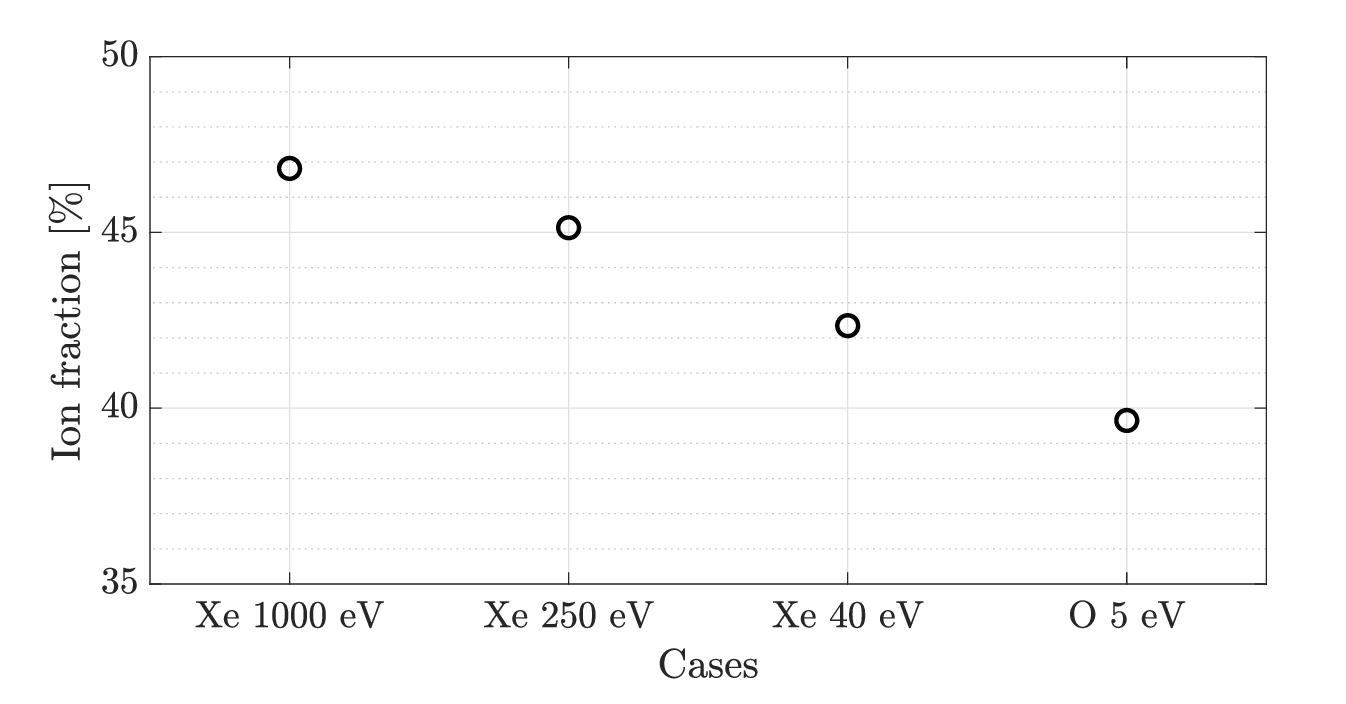}
  \caption{Ion fraction of the outlet flux}
  \label{fig:DoI}
\end{figure}


Fig.~\ref{fig:flux} shows the total outlet flux, combining ions, neutrals, and sputtered atoms. In the Xe 1000 eV case, the total flux is $5.28\times10^{19}$ m$^{-2}$s$^{-1}$, consisting of 46.7\% ion flow, 49.0\% neutral flow, and 4.2\% sputtered flow. When the ion energy is reduced to 250 eV and 40 eV, the sputtered fraction decreases to 1.2\% and 0.04\%, respectively. This indicates that lowering the beam ion energy below 100 eV may be necessary to suppress sputtering on the intake device surface, thereby minimizing its influence on $\eta_c$ evaluation. As the beam ion energy decreases, the ion fraction within the total outlet flux also decreases, as shown in Fig.~\ref{fig:DoI}. Moreover, the O 5 eV case exhibits an even lower ion fraction than the Xe 40 eV case. At low ion energy and mass, enhanced beam divergence reduces the probability of ions transmitting directly through the intake device to the outlet without ISI. Even though more than 86\% of the inlet flow initially consists of ions, a substantial fraction is neutralized, resulting in a neutral-dominated flow at the outlet. Accordingly, measurement of neutral particles is essential for accurately evaluating $\eta_c$, in addition to ion diagnostics. Table~\ref{tab:capture_eff} summarizes the estimated $\eta_c$ obtained using different measurement approaches. When both ion and neutral atom flow rates are considered, the $\eta_c$ of the parabolic intake device employed in this study is expected to be 7.14--8.23\%, depending on the thruster exhaust conditions. However, if only plasma diagnostics are applied, $\eta_c$ would be underestimated, yielding values in the range of 3.13--4.12\%. Neutral-only measurement also would result biased $\eta_c$ to 4.19--4.31\%. These results indicate that simultaneous measurement of both ion and neutral flows is essential for reliable $\eta_c$ evaluation in stand-alone intake-device testing using an EP plasma plume within conventional vacuum chambers.

\begin{table}[h]
  \centering
  \caption{Inlet/outlet flux and $\eta_c$ for different cases}
  \label{tab:capture_eff}
  \begin{tabular}{lcccc}
    \toprule
    \textbf{Properties} & \textbf{Xe 1000 eV} & \textbf{Xe 250 eV} & \textbf{Xe 40 eV} & \textbf{O 5 eV} \\
    \midrule
    Inlet ion flux [$\mathrm{m^{-2}\,s^{-1}}$]   & $3.96\times10^{19}$ & $3.37\times10^{19}$ & $1.65\times10^{19}$ & $7.56\times10^{18}$ \\
    Inlet neutral atom flux [$\mathrm{m^{-2}\,s^{-1}}$] & $1.93\times10^{18}$ & $1.97\times10^{18}$ & $1.64\times10^{18}$ & $1.14\times10^{18}$ \\
    Outlet ion flux [$\mathrm{m^{-2}\,s^{-1}}$]  & $2.47\times10^{19}$ & $1.86\times10^{19}$ & $8.44\times10^{18}$ & $3.65\times10^{18}$ \\
    Outlet neutral atom flux [$\mathrm{m^{-2}\,s^{-1}}$] & $2.59\times10^{19}$ & $2.21\times10^{19}$ & $1.15\times10^{19}$ & $5.55\times10^{18}$ \\
    Outlet sputtered atom flux [$\mathrm{m^{-2}\,s^{-1}}$] & $2.24\times10^{18}$ & $5.12\times10^{17}$ & $8.14\times10^{15}$ & - \\

    $\eta_c$ (ion-only) [\%]   & 4.12 & 3.52 & 3.13 & 2.83 \\
    $\eta_c$ (neutral-only) [\%]& 4.21 & 4.19 & 4.26 & 4.31 \\
    $\eta_c$ (Total) [\%]       & 8.23 & 7.71 & 7.39 & 7.14 \\
    \bottomrule
  \end{tabular}
\end{table}

\section{Conclusion}


This work numerically investigates how EP plasma plume–intake interactions govern measurable capture efficiency $\eta_c$ in vacuum facilities. Plume ion energy and species mass influence plume divergence and ion neutralization, yielding neutral-dominated outlet flows. High energy of plume ions may increase wall sputtering that can bias the measurements. Reliable $\eta_c$ evaluation requires simultaneous ion and neutral diagnostics and operation at plume ion energies below 100 eV to limit sputter-induced error. Under these conditions, stand-alone intake tests in conventional vacuum chambers may provide a practical basis for intake performance evaluation and ABEP feasibility analysis. Because $\eta_c$ is sensitive to surface treatment and coating materials, future work will examine GSI/ISI variations in ground tests~\cite{Moon2023vac,Huh2025,Huh2024rgd,Murray2017,Xu2023,park2025}. Moreover, similarity between ground-test and on-orbit performance will be addressed to establish the validity of ground-based measurements.

\section*{Acknowledgments}

This work was supported by the Agency For Defense Development by the Korean Government (UG233092TD). This work was supported by the National Supercomputing Center with supercomputing resources including technical support(KSC-2025-CRE-0427).


\bibliography{mybibfile}

\end{document}